\documentclass[onecolumn,nobibnotes,nofootinbib,superscriptaddress,11pt]{revtex4}
\usepackage{amsmath,amssymb,bm}
\usepackage[a4paper,bindingoffset=0.2in,left=0.8in,right=0.8in,top=1in,bottom=1in,footskip=.25in]{geometry}
\usepackage{graphicx,subfigure,epsfig}
\usepackage{bigints}
\usepackage{color}
\usepackage[breaklinks,colorlinks,urlcolor=blue,citecolor=blue,linkcolor=blue]{hyperref}
\usepackage{mciteplus}
\definecolor{lcolor}{rgb}{0.5,0,0}
\definecolor{citcolor}{rgb}{0,0.3,0.0}
\usepackage[capitalise]{cleveref}
\usepackage{todonotes}

\newcommand \footnoteONLYtext[1]
{
	\let \mybackup \thefootnote
	\let \thefootnote \relax
	\footnotetext{#1}
	\let \thefootnote \mybackup
	\let \mybackup \imareallyundefinedcommand
}

\newcommand{\be}{\begin{equation}}
\newcommand{\ee}{\end{equation}}
\newcommand{\beq}{\begin{eqnarray}}
\newcommand{\eeq}{\end{eqnarray}}
\newcommand{\benn}{\begin{displaymath}}
\newcommand{\eenn}{\end{displaymath}}
\newcommand{\beann}{\begin{eqnarray*}}
\newcommand{\eeann}{\end{eqnarray*}}

\begin{document}

\title{New exact analytical solution of the nonlinear Gribov-Levin-Ryskin-Mueller-Qiu equation}
\author{Yanbing Cai}
%\email{myparticle@163.com}
\affiliation{Guizhou Key Laboratory in Physics and Related Areas, Guizhou University of Finance and Economics, Guiyang 550025, China}
\affiliation{Southern Center for Nuclear-Science Theory (SCNT), Institute of Modern Physics, Chinese Academy of Sciences, Huizhou 516000, China}
\author{Xiaopeng Wang}
\footnoteONLYtext{These authors contributed equally: Yanbing Cai and Xiaopeng Wang.}
\affiliation{Institute of Modern Physics, Chinese Academy of Sciences, Lanzhou 730000, China}
\affiliation{School of Nuclear Science and Technology, Lanzhou University, Lanzhou 730000, China}
\affiliation{School of Nuclear Science and Technology, University of Chinese Academy of Sciences, Beijing 100049, China}
\author{Xurong Chen}
%\email{xchen@impcas.ac.cn}
\email{xchen@impcas.ac.cn (Corresponding author)}
\affiliation{Institute of Modern Physics, Chinese Academy of Sciences, Lanzhou 730000, China}
%\affiliation{Guangdong Provincial Key Laboratory of Nuclear Science, Institute of Quantum Matter, South China Normal University, Guangzhou 510006, China}
\affiliation{School of Nuclear Science and Technology, University of Chinese Academy of Sciences, Beijing 100049, China}

%\pacs{}

\begin{abstract}
The GLR-MQ equation is a nonlinear evolution equation that takes into account the shadowing effect, which tames the growth of the gluon at small-$x$. In this study, we analytically solve for the first time the nonlinear GLR-MQ equation using the homogeneous balance method. The definite solution of the GLR-MQ equation is obtained by fitting the MSTW2008LO gluon distribution data. We find that the geometric scaling is an intrinsic property of our analytical solution and the gluon distribution functions from our solution are able to reproduce the MSTW2008LO data. These results indicate that our analytical solution from the homogeneous balance method is valid to describe the gluon behavior at small-$x$. Moreover, the saturation scale $Q_s$ has been extracted from our analytical solution, we find that the energy-dependent saturation scale obeys the exponential law $Q_s^2\,\propto\,Q_0^2 e^{\lambda Y}$.
\end{abstract}

\maketitle

%-----------------------------------------------------------------------------
\section{Introduction}
\label{intro}
The collinear factorization theorem is one of the most successful tools to describe phenomena in high-energy processes. In the framework of collinear factorization theorem, the parton distribution functions (PDFs), which are universal and process independent, are key ingredients to describe the cross sections or scattering amplitudes. The PDFs are expressed as functions of $(x,Q^2)$ with $x$ the parton momentum fraction (Bjorken-$x$) and $Q^2$ the virtuality of the photon. In perturbative quantum chromodynamics (pQCD), the $Q^2$ dependence of PDFs are usually given by the linear Dokshitzer-Gribov-Lipatov-Altarelli-Parisi (DGLAP) evolution equations \cite{Altarelli:1977zs,Gribov:1972ri,Lipatov:1974qm,Dokshitzer:1977sg}. The DGLAP equations are derived in the limit of large $x$ and $Q^2$. When one goes into the small-$x$ region, the partons densities obtained from DGLAP evolution equations increase steeply as the parton split. This property of increasing parton distribution functions has been observed in electron-proton scattering at HERA \cite{H1:1993jmo,ZEUS:1997etp}.

However, the gluon density can not increase infinitely due to the Froissart-Martin bound \cite {Froissart:1961ux,Martin:1962rt} and unitarity \cite{kovchegov_levin_2012}.
According to the uncertainty principle, the resolution scale in transverse plane is of order $1/Q^2$. In extremely small-$x$ at fixed $Q^2$, the gluon density becomes particularly dense. As a result, the gluons start to overlap in the transverse space. In this region, the correlative interactions between gluons become important. In other words, the probability of recombining two gluons into one increases, which will tame the increase of gluons and eventually lead to the formation of gluon saturation \cite{Albacete:2014fwa,Blaizot:2016qgz}. This saturation phenomenon is widely studied by the nonlinear evolution equations, such as the Jalilian-Marian-Iancu-McLerran-Weigert-Leonidov-Kovner (JIMWLK) equation \cite{Jalilian-Marian:1996mkd,Jalilian-Marian:1997qno,Jalilian-Marian:1997jhx,Iancu:2001md,Ferreiro:2001qy,Iancu:2001ad,Iancu:2000hn} and its mean-field version known as the Balitsky-Kovchegov (BK) equation \cite{Balitsky:1995ub,Kovchegov:1999yj}, the Gribov-Levin-Ryskin-Mueller-Qiu (GLR-MQ) equation \cite{Gribov:1983ivg,Mueller:1985wy}, and the Modified-DGLAP equation \cite{Zhu:2003xta}. The studies related to these nonlinear equations have considerably improved our understanding of gluon saturation.

The GLR-MQ equation is calculated from the pQCD approach by taking into account the gluon ladders recombination by Mueller and Qiu \cite{Gribov:1983ivg,Mueller:1985wy}. It can be viewed as an updated version of the linear DGLAP equation with nonlinear recombination corrections. Therefore, the study of GLR-MQ equation is useful for a comprehensive understanding of gluon recombination and saturation. Recently, the GLR-MQ equation has been studied numerically and analytically. In the aspect of numerical study, the GLR-MQ equation has been used to investigate the nonlinear corrections for the singlet and gluon distributions  \cite{Rausch:2022nkv} and to evaluate the nuclear parton distributions \cite{Boroun:2023clh}. These numerical studies indicate that the nonlinear corrections are significant for heavy nuclei. In the aspect of analytical study, the GLR-MQ equation has been analyzed by using the semi-classical approximation \cite{He:1999wh} and Regge-like ansatz \cite{Devee:2014ida,Devee:2014fna,Lalung:2017omk,Phukan:2017lzp}. However, the approximations used in these approaches have limitations. Such as, the Regge behavior of the gluon distribution was pre-assumed in Regge-like method while it is only valid at small-$x$ and at intermediate $Q^2$ \cite{Ball:1994du,Kotikov:1995ny}. In this study, we shall analytically solve the nonlinear GLR-MQ equation using the homogeneous balance method. The homogeneous balance method is a mathematical approach to find an analytical solution of the nonlinear partial equations \cite{WANG1996279}. By fitting the MSTW2008LO gluon distribution data, we obtain the definite solution of the GLR-MQ equation. The numerical results demonstrate that our analytical solution from the homogeneous balance method is valid to describe the gluon behavior at small-$x$, in which the geometric scaling is naturally satisfied.

\section{The nonlinear GLR-MQ equation and its analytical solution}

In this section we first briefly introduce the nonlinear GLR-MQ equation. Then, we provide a detailed description of the process to obtain the analytical solution of the GLR-MQ equation using the homogeneous balance method.

As we know, the microscopic structure of proton is described by the partons (quark and gluon) distributions. In small-$x$ region ($x < 0.01$), these partons are mostly occupied by gluons. The number density of gluon in a proton can be evaluated by the DGLAP evolution equation \cite{kovchegov_levin_2012}
 \be
 \label{g_DGLAP}
 Q^{2}\frac{\partial g(x,Q^{2})}{\partial Q^{2}}= \frac{\alpha_{s}}{2 \pi}\int_x^1 \frac{dz}{z}P_{gg}(z)g(\frac{x}{z},Q^{2}) \ ,
\ee
where $P_{gg}$ is the gluon-gluon splitting function. Note that a full evolution of the gluon density should include the contributions from quark distribution. However, in the small-$x$ region, quark contributions can be neglected as the gluon is dominant and its number density grows much faster than that of quark. Therefore, in contrast to standard DGLAP, only the gluon-gluon splitting function and gluon density are included in the right hand side of Eq.(\ref{g_DGLAP}). What is more, at small-$x$ the gluon-gluon splitting function $P_{gg}(z)$ is proportional to $2N_{c}/z$, with $N_{c}$ the number of color. Applying this approximation, Eq.(\ref{g_DGLAP}) can be written as a differential equation \cite{kovchegov_levin_2012}
 \be
 \label{g_DGLAP_DLA}
 \frac{\partial^{2} G(x,Q^{2})} {\partial Y \partial\ln\frac{Q^{2}}{\Lambda^{2}}}= \frac{\alpha_{s}N_{c}}{\pi}G(x,Q^{2}) \ ,
\ee
where $Y=\ln (1/x)$ is the rapidity, $\Lambda$ is the initial-virtulity scale, and $G(x,Q^{2})=xg(x,Q^{2})$ is the gluon distribution function. From Eq.(\ref{g_DGLAP_DLA}), we can clearly see that the approximation is equal to the resummation of series of two logarithmic $\alpha_{s}\ln (1/x)\ln(Q^2/\Lambda^2)$, which is called double logarithmic approximation (DLA).

In DLA, the gluon distribution function rises steeply as a power of $x$ at small-$x$ region. However, according to the Froissart-Martin bound \cite {Froissart:1961ux,Martin:1962rt}, the total cross section in QCD cannot grow faster than $\ln s$ ($s$ is the energy squared). In contrast, the cross section in DLA growth much faster than any power of $\ln s$, which means that the DGLAP evolution violates unitarity. This problem also arises when evaluating the gluon density via BFKL evolution, since the BFKL in DLA is equivalent to the DGLAP in DLA. To tackle the unitarity problem, Gribov, Levin, and Ryskin propose that for a dense system multiple BFKL ladder becomes important \cite {Gribov:1983ivg}. Due to the high gluon density, the emergence of two ladders into one ladder will be possible. Taking into account this recombination effect brings a quadratic correction term to the linear BFKL equation and leads to the Gribov-Levin-Ryskin (GLR) equation. In fact, at extremely small values of $x$, gluons overlap spatially as its density becomes large enough, the recombination becomes very important. Mueller and Qiu reconsider the recombination of gluon ladders in DLA and obtain a new evolution equation, the well-known GLR-MQ equation \cite{Mueller:1985wy}
 \be
 \label{g_GLR-MQ_DLA}
 \frac{\partial^{2} G(x,Q^{2})} {\partial Y \partial\ln \frac{Q^{2}}{\Lambda^{2}}}= \frac{\alpha_{s}N_{c}}{\pi}G(x,Q^{2})-\frac{\alpha_{s}^{2}\gamma}{Q^{2}R^{2}}G^{2}(x,Q^{2}) \ ,
\ee
where $\gamma=81/16$ for $N_{c}=3$, $R$ is the correlation radius between two interacting gluons.

For Eq.(\ref{g_GLR-MQ_DLA}), there are two main features. First, the first term in the right hand side is the linear DGLAP equation in the DLA. This term leads to the growth of the number of gluons. Second, the second term in the right hand side is a nonlinear term corresponding to the gluon recombination effect. The nonlinear recombination correction has a minus sign, which decreases the number of gluons. So, the nonlinear term has shadowing effect. The shadowing effect tames the growth of gluons, which can be viewed as a precursor of the gluon saturation \cite{Gelis:2010nm}. The nonlinear term in the GLR-MQ equation becomes significant only near the boundary of the saturation region. Therefore, this equation is valid only near the boundary of the critical line \cite{Eskola:2002yc,AyalaFilho:1998rh}.

The behavior of the gluon distribution from the GLR-MQ evolution equation is extremely critical to understand the saturation in small-$x$. However, the GLR-MQ evolution equation is a nonlinear equation with a quadratic term.  It is difficult to obtain a solution directly. In the literature, this equation was evaluated numerically \cite{Rausch:2022nkv,Boroun:2023clh} or under some approximation \cite{He:1999wh,Devee:2014ida,Devee:2014fna,Lalung:2017omk,Phukan:2017lzp}. In this study, we shall use the homogeneous balance method. Homogeneous balance method is a useful approach to find the solution of a nonlinear partial equations $P\left(u, u_{x}, u_{t}, u_{x x}, u_{x t}, u_{t t}, \ldots\right)=0$ \cite{WANG1996279,WANG1995169,WANG199667,FAN1998389}, where the subscripts denote the derivatives. In this approach, one should first choose a suitable linear combination of the heuristic solutions, which in general contains undetermined coefficients. If the heuristic solution is indeed a solution of a nonlinear partial equation. Then one can determine the coefficients step by step \cite{WANG1996279}. The main principle in homogeneous balance method is that the highest nonlinear term and the highest order partial derivative term should be balanced \cite{WANG1996279}.  The homogeneous balance method has been widely used to search the solutions of numerous nonlinear evolution equations in mathematical physics, such as the KdV equation \cite{FAN1998389}, the FKPP equation \cite{Yang:2020jmt}, and the Balitsky-Kovchegov equation \cite{Wang:2020stj,Cai:2023iza}.

For mathematical convenience, we first simplify Eq.(\ref{g_GLR-MQ_DLA}) through altering the variables. Let $t=Y$, $z=\ln(Q^2/\Lambda^2)$, and $G=e^{z}u$, then Eq.(\ref{g_GLR-MQ_DLA}) becomes
\be
\label{GLR-MQ_sim}
u_{tz}+u_t = \frac{\alpha_s N_c}{\pi}u-\frac{\alpha^2_s  \gamma}{ R^2 \Lambda^2}u^2 \ .
\ee
As mentioned above, the GLR-MQ evolution equation is valid in the border of the critical line. In this region, $Q_{s} \gg \Lambda_{QCD}$, the value of QCD coupling is small. Therefore, $\alpha_s$ can be viewed as a fixed constant as assumed in Ref.\cite{Laenen:1995fh}. Letting $\alpha_s N_c/\pi=\alpha$, $\alpha^2_s  \gamma/ (R^2 \Lambda^2)=\beta$, Eq.(\ref{GLR-MQ_sim}) has the following form
\begin{equation}
\label{GLR-MQ_fin}
u_t+u_{tz}-\alpha u+\beta u^2=0 \ .
\end{equation}

Eq.(\ref{GLR-MQ_fin}) is a partial differential equation with constant coefficients, it can be solved using the homogeneous balance method. In order to obtain the exact solution, we assume that Eq.(\ref{GLR-MQ_fin}) has a heuristic solution as \cite{FAN1998389}
\be
\label{hs}
u(t, z)=\sum_{m+n=1}^{N} a_{m+n} \frac{\partial^{(m+n)}}{\partial t^{m} \partial z^{n}} f(\varphi(t, z)) \ .
\ee
Submit Eq.(\ref{hs}) into Eq.(\ref{GLR-MQ_fin}), the highest nonlinear term is written by
\be
\label{nonlinear_t}
u^{2}=f^{(N)} f^{(N)} \varphi_{t}^{2 m} \varphi_{z}^{2 n}+\cdots \ .
\ee
And the highest order partial derivative term is written by
\be
\label{derivative_t}
u_{tz}=f^{(N+2)}\varphi^{m+1}_t \varphi^{n+1}_z+\cdots \ .
\ee
According to the homogeneous balance principle, the power of $\partial/\partial t$ and $\partial/\partial z$  in Eq.(\ref{nonlinear_t}) and Eq.(\ref{derivative_t}) should be balance, respectively \cite{WANG1996279}. Based on this principle, we can obtain
\be
\label{hs_cof}
m=1\ , ~~~~~~~
n=1\ , ~~~~~~~\mathrm{and}~~
N=m+n=2\ .
\ee
Make $a_{m+n}=1$, then the heuristic solution becomes
\be
u(t,z)=f_{tz}+a_1 f_z + a_2 f_t +b \ .
\ee
Substitute this heuristic solution into equation(\ref{GLR-MQ_fin}), and make the highest power of the derivative of $\varphi$ to be zero, then we can get
\be
\label{par_solution}
f^{(4)}+\beta(f'')^2=0  \ .
\ee
The particular solution for Eq.(\ref{par_solution}) has the form
\be
\label{par_solutionf}
f=\frac{6}{\beta}\ln\varphi\ .
\ee

Suppose the formalism of $\varphi$ has the traveling wave structure
\be
\label{phi}
\varphi=h+e^{k z+ c t + \theta} \ .
\ee
where $h$, $k$ and $c$ are constants to be determined.
Substitute Eq.(\ref{phi}) into Eq.(\ref{GLR-MQ_fin}), collect all terms with the same order of derivative of $f$, and set the coefficients of each order of derivative to be zero, then we can get a set of algebra equations with the undetermined parameters. Solving the algebraic equations, we can obtain the values of the coefficients and parameters.
\be
\label{u_cof}
a_{2}=\frac{5\alpha + 6a_{1}}{25\beta} \ , ~~~~~~~
k=\frac{1}{5} \ ,~~~~~~~
c=\frac{-5\alpha}{6} \ , ~~~~~~~\mathrm{and}~~
b=\frac{\alpha}{\beta} \ .
\ee

Submitting Eqs.(\ref{par_solutionf}) and (\ref{u_cof}) into Eq.(\ref{GLR-MQ_fin}), the solution is given by
\be
\label{us_f}
u(t, z)=\frac{\alpha}{\beta}\left[1-\frac{1}{1+h e^{\left(\frac{5\alpha t}{6}-\frac{z}{5}\right)}}\right]^2 \ .
\ee
The solution of the GLR-MQ evolution equation with fixed coupling constant is given by
\be
\label{Gs_f}
G(x,Q^2)=\frac{\alpha}{\beta}\frac{Q^2}{\Lambda^2}\left[1-\frac{1}{1+{h}(\frac{Q^2}{\Lambda^2})^{\frac{1}{5}}x^\frac{5\alpha }{6}}\right]^2 \ ,
\ee
where $h$ is still a free independent parameter, which will be determined by fitting experimental data. Eq.(\ref{Gs_f}) has surprising outcomes related to geometric scaling. In high energy (small-$x$) region, the saturation model predicts that the inclusive and diffractive deep inelastic data have geometric scaling \cite{Stasto:2000er}. This phenomenon holds for BK equation \cite{Munier:2003vc}. By employing the parameterization of proton structure function, the investigation for the GLR-MQ equation in LO approximated model indicates that the GLR-MQ equation also has geometric scaling \cite{Boroun:2022akh}. Moreover, the geometric scaling of the dipole cross section corresponds to the geometric scaling of $\alpha_s G/Q^2$ \cite{Boroun:2015fna}. For better seeing this, Eq.(\ref{Gs_f}) can be rewritten as
\be
\label{Gs_gs}
G(x,Q^2)=\frac{\alpha}{\beta}\frac{Q^2}{\Lambda^2}\left[1-\frac{1}{1+{h}(\frac{Q_s^2}{Q^2})^{\frac{1}{5}}}\right]^2 \ ,
\ee
%------------------------------------------------------------------------------------------------------------------------------------------------------------
where
\be
\label{Qs_gs}
Q_s^2=\Lambda^2 x^\frac{-25\alpha }{6}.
\ee
Now $\alpha_s G/Q^2$ depend on a single variable $Q_s^2/Q^2$. Therefore, the geometric scaling is naturally satisfied in our solution.
%-------------------------------------------------------------------------------------------------------------------------------------------------------------
\section{Numerical results}
\label{results}

The definitive solution for GLR-MQ evolution equation with fixed coupling constant is obtained by fitting the gluon distribution function from MSTW2008LO \cite{Martin:2009iq}. In our simulation the correlation radius $R$ is set to be $2~\text{GeV}^{-1}$ and $5~\text{GeV}^{-1}$. The kinematic range is $10^{-4} \leq x \leq 10^{-2}$ and $5~ \text{GeV}^2 \leq Q^{2}\leq 20~\text{GeV}^2$. This kinematic range is divided into $50*50$ grids (logarithm in $x$ and $Q^{2}$). The resulting parameters and $\chi^{2}$ per number of points ($\chi^{2}/\text{n.o.p.}$) are shown in Table ~\ref{table:1}. From the $\chi^{2}/\text{n.o.p.}$, we can see that our definitive solution is consistent with the gluon distribution function from MSTW2008LO. We have checked that the parameters and $\chi^{2}/\text{n.o.p.}$ are insensitive to the details of the number of grids. It should be noted that $R = 5~\text{GeV}^{-1}$ corresponds to the radius of proton, whereas $R = 2~\text{GeV}^{-1}$ corresponds to the radius of hot spot in a proton. The $\chi^{2}/\text{n.o.p.}$ in $R = 2~\text{GeV}^{-1}$ is better than that in $R = 5~\text{GeV}^{-1}$. This implies that the behavior of the gluon distribution favors concentration in the hot spot.

\begin{table}[h!]
  \begin{center}
  \caption{Parameters and $\chi^{2}/\text{n.o.p.}$ results from the fit to MSTW2008LO gluon distribution data.}
  \begin{tabular}{ccccccccc}
  \hline
    &  &  $\alpha_{s}$ & $\Lambda~(\text{GeV})$ &  $h$ & $\chi^{2}/\text{n.o.p.}$ \\
  \hline
    & $R = 2~\text{GeV}^{-1}$   & 0.165  & 0.209  & -0.406   & 1.544  \\
    & $R = 5~\text{GeV}^{-1}$   & 0.205  & 0.224  & -0.174   & 1.983  \\
  \hline
    \end{tabular}%
  \label{table:1}%\vspace{-1mm}
  \end{center}
\end{table}

\begin{figure}[h!]
\begin{center}
\includegraphics[width=0.47\textwidth]{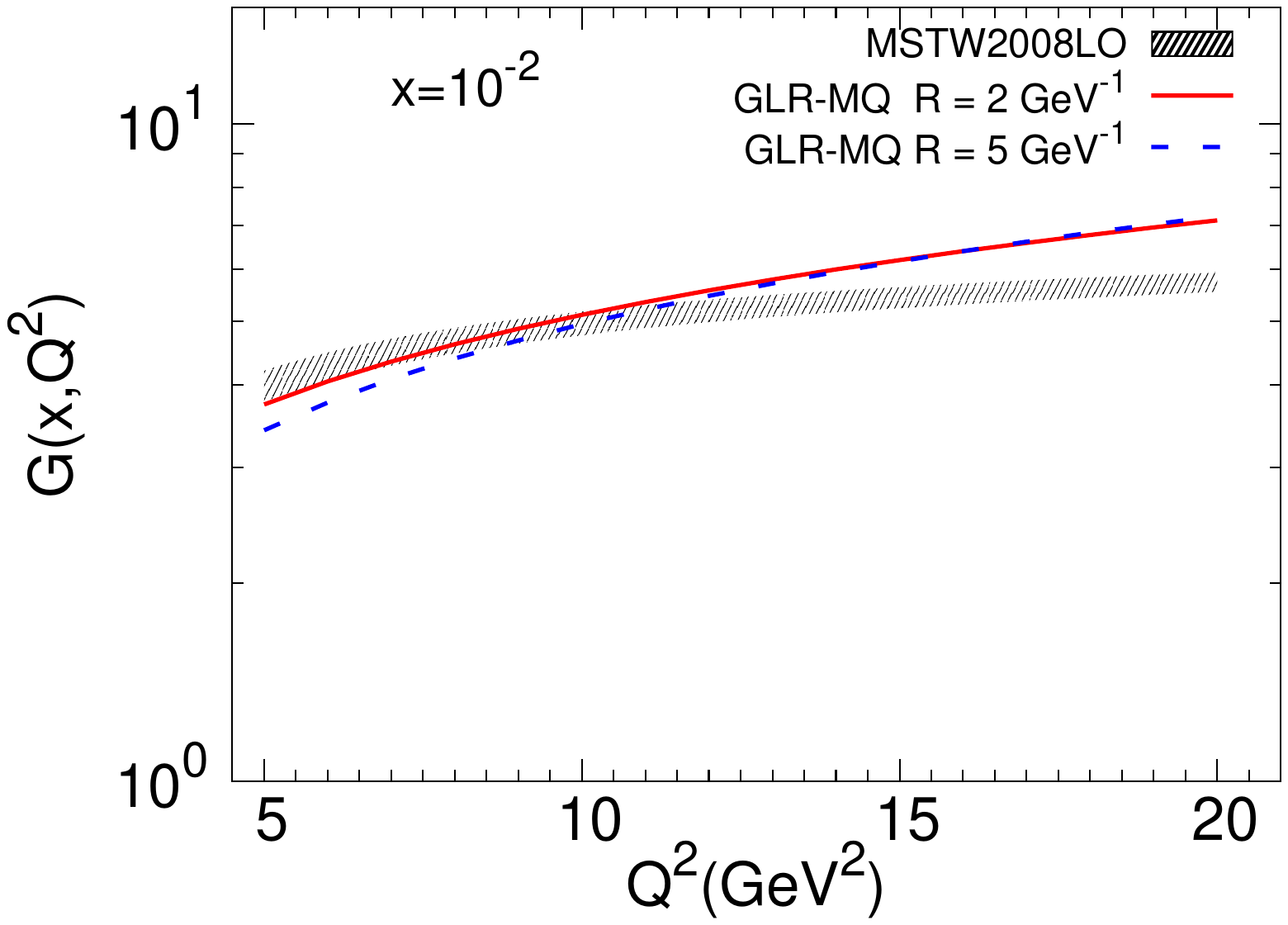}
\hspace{-0.3cm}
\includegraphics[width=0.47\textwidth]{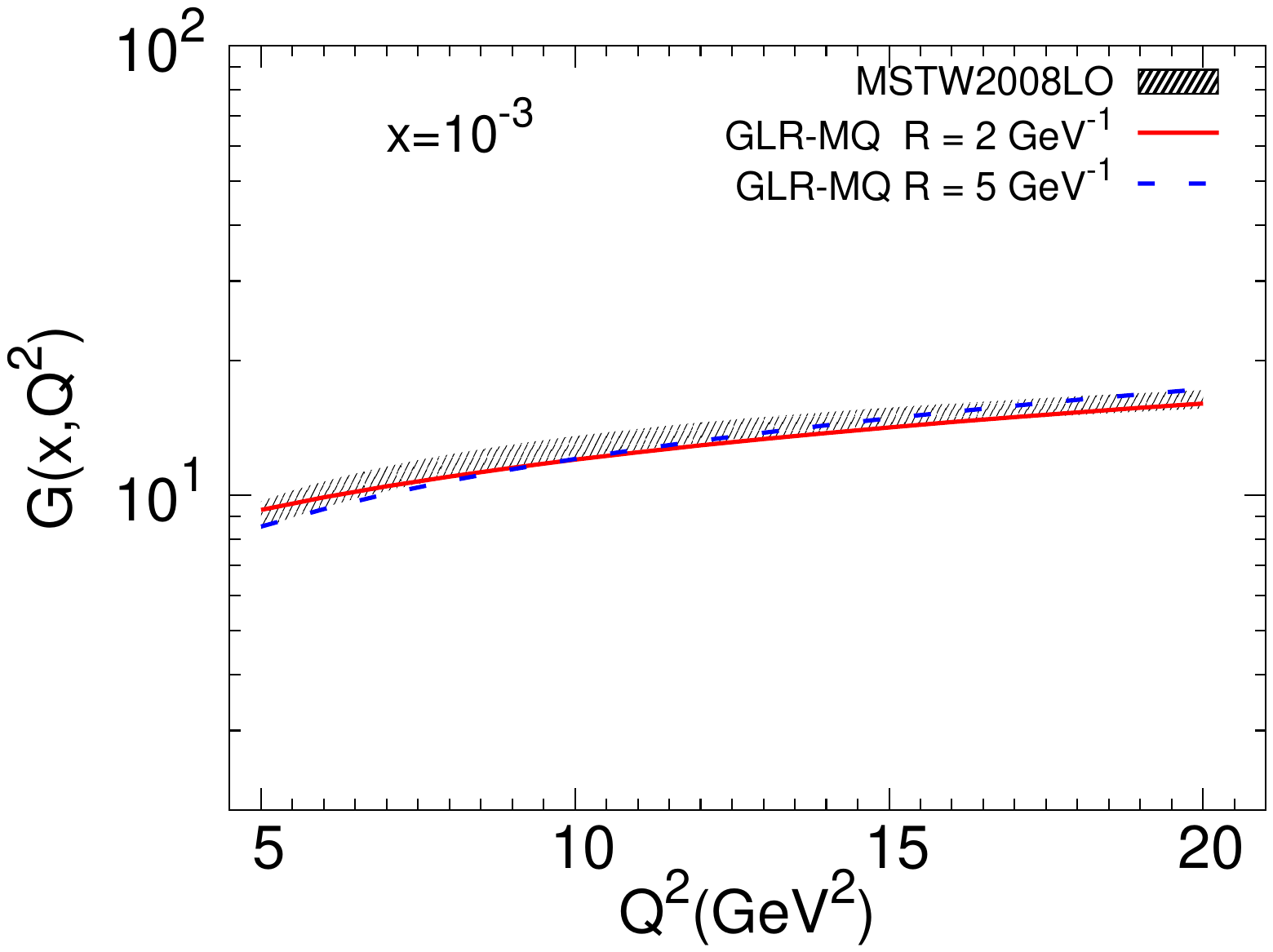}
%\vspace{0cm}\hspace{0cm}
\includegraphics[width=0.47\textwidth]{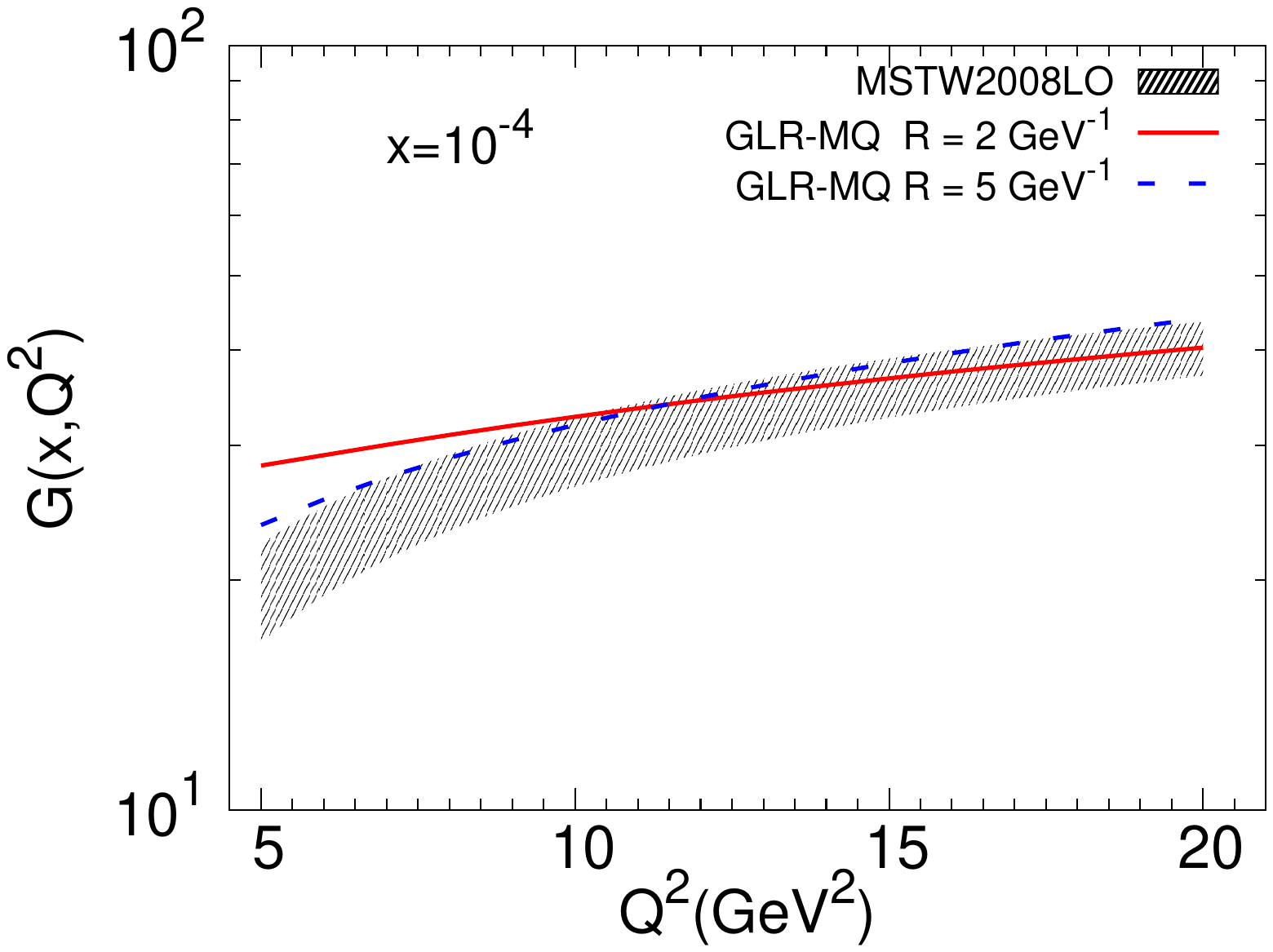}
%\vspace{0cm}\hspace{-0.3cm}
%%%%%%%%%%%%%%%%%%%%%%%%%%%%%%%%%%%%%%%%%%%%%%%%%%
\end{center}
\vspace{-0.5cm}
\caption{The gluon distribution functions $G(x,Q^2)$ as a function of virtuality $Q^{2}$ at different Bjorken-$x$. The solid lines represent the results from $R = 2~\text{GeV}^{-1}$, the dashed lines represent the results from $R = 5~\text{GeV}^{-1}$, and the bands represent the data from MSTW2008LO with error bars.}
\label{fig:G_Q2}
\end{figure}

\begin{figure}[h!]
\begin{center}
\includegraphics[width=0.47\textwidth]{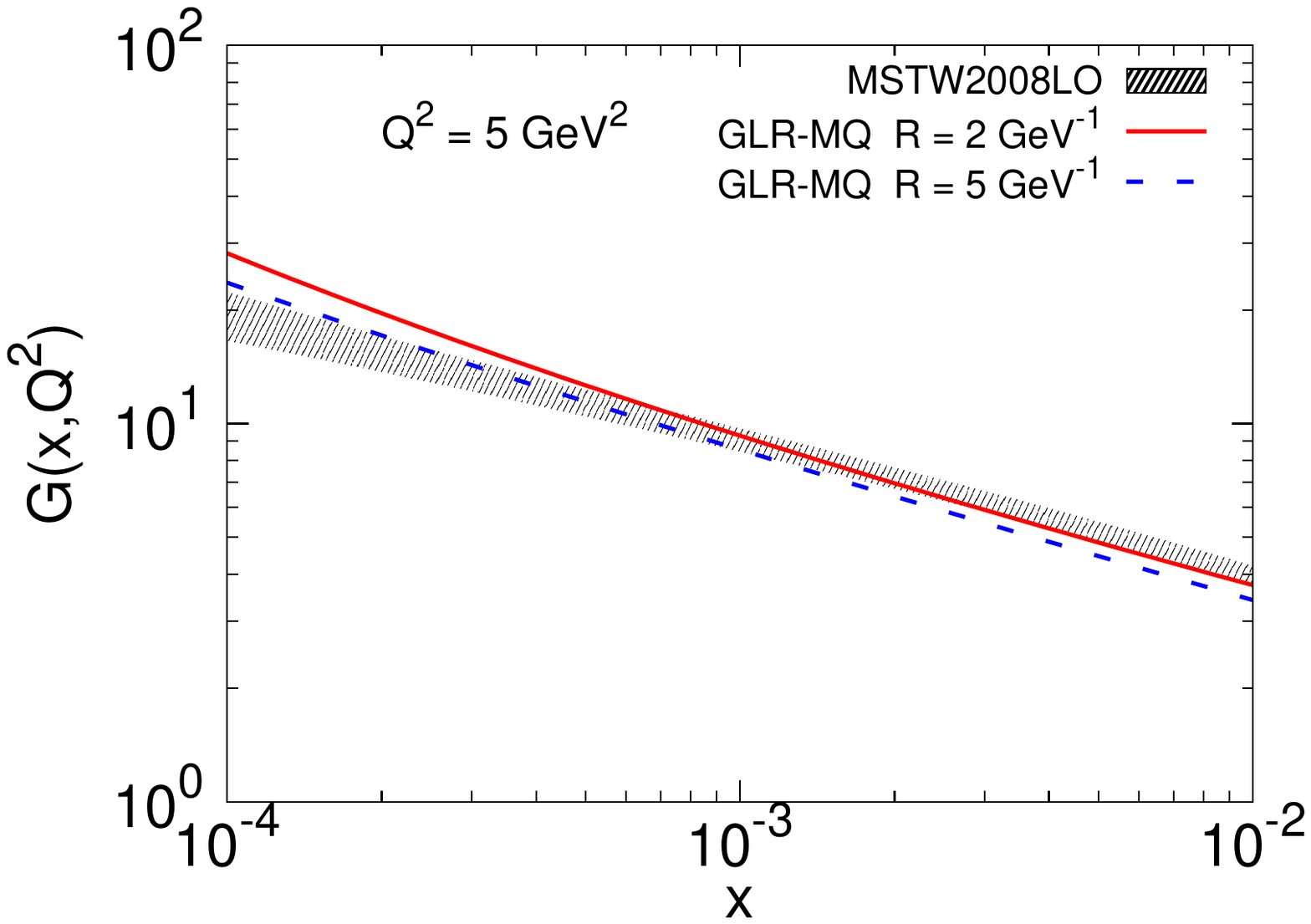}
\hspace{-0.3cm}
\includegraphics[width=0.47\textwidth]{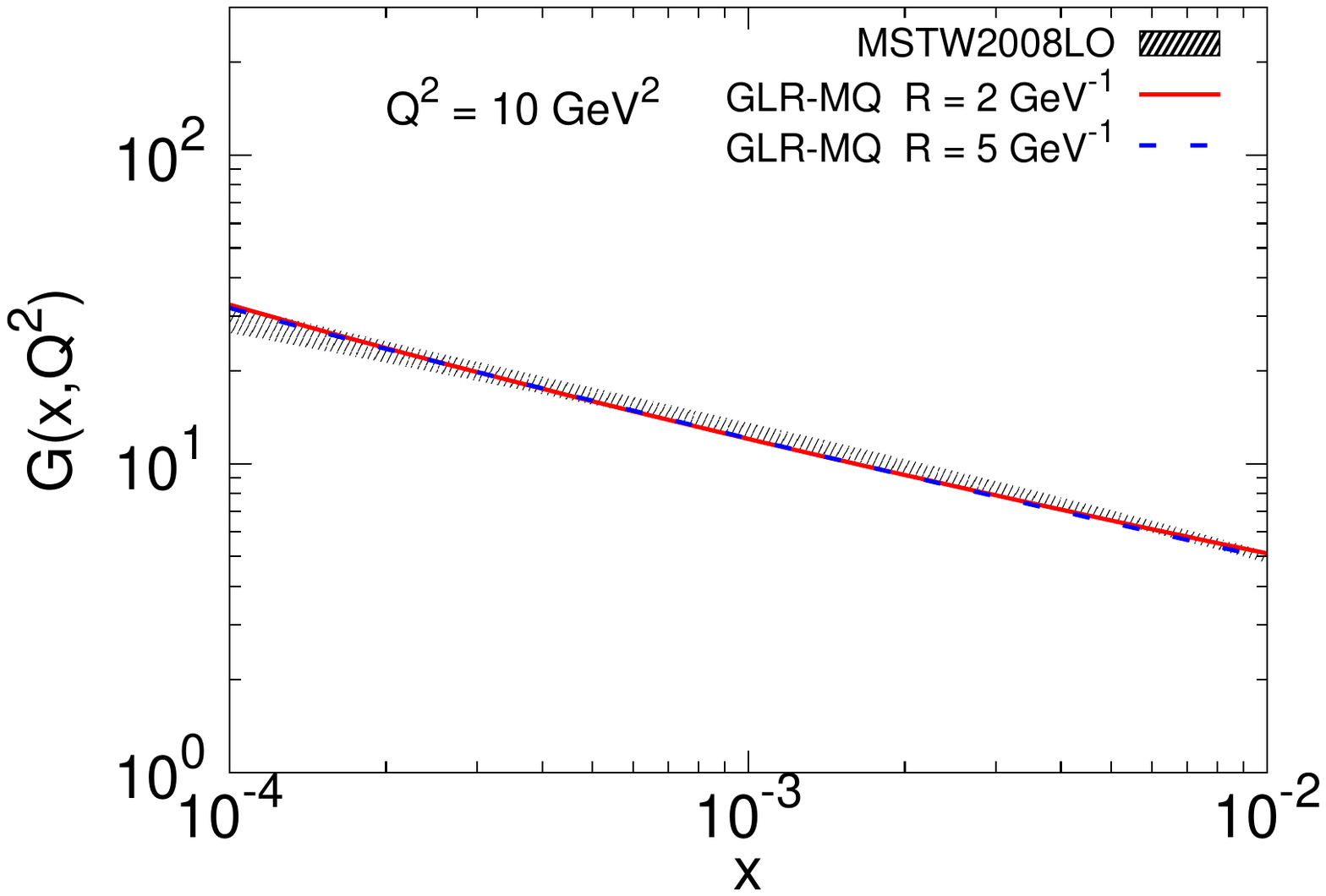}
%\vspace{0cm}\hspace{0cm}
\includegraphics[width=0.47\textwidth]{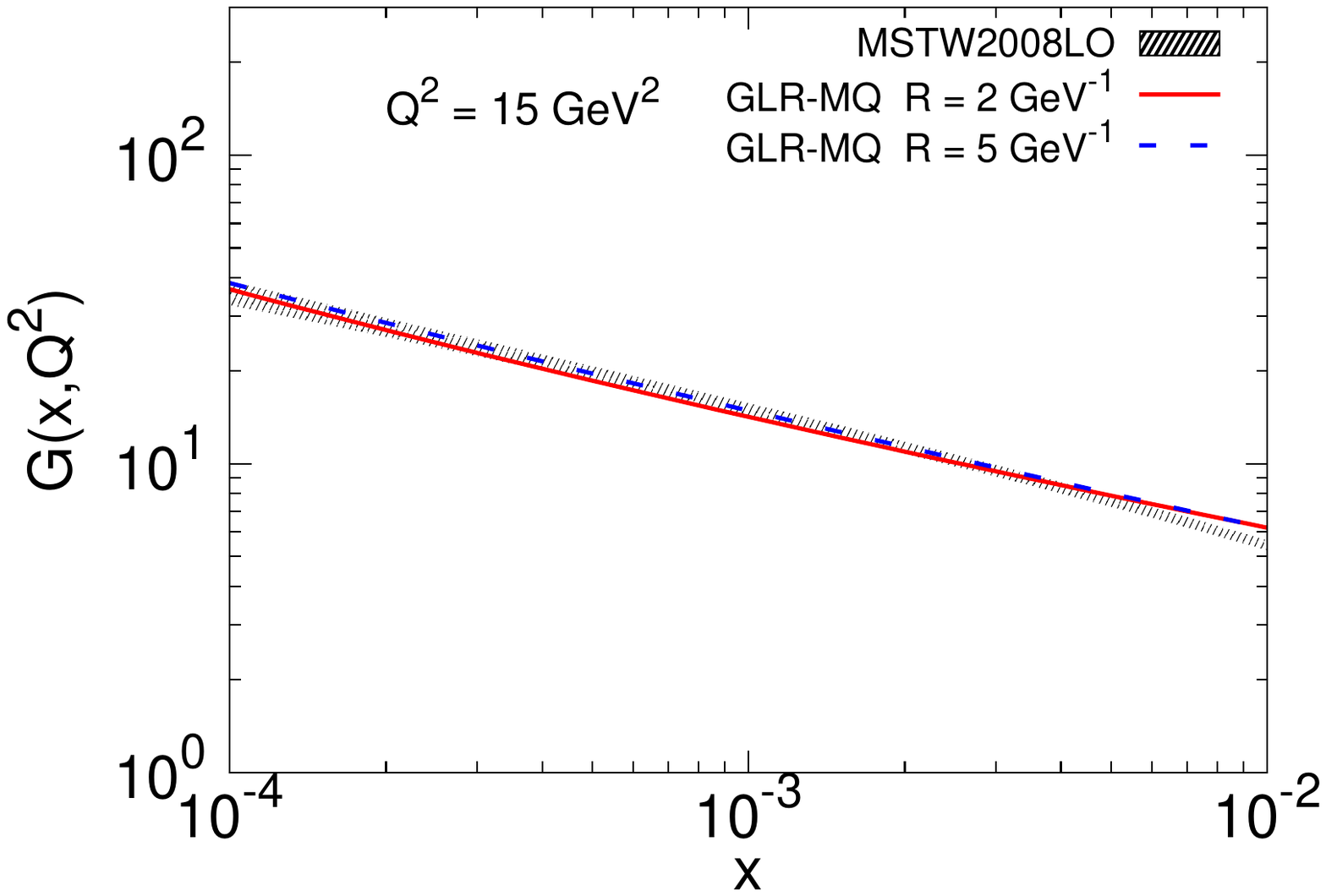}
\hspace{-0.3cm}
\includegraphics[width=0.47\textwidth]{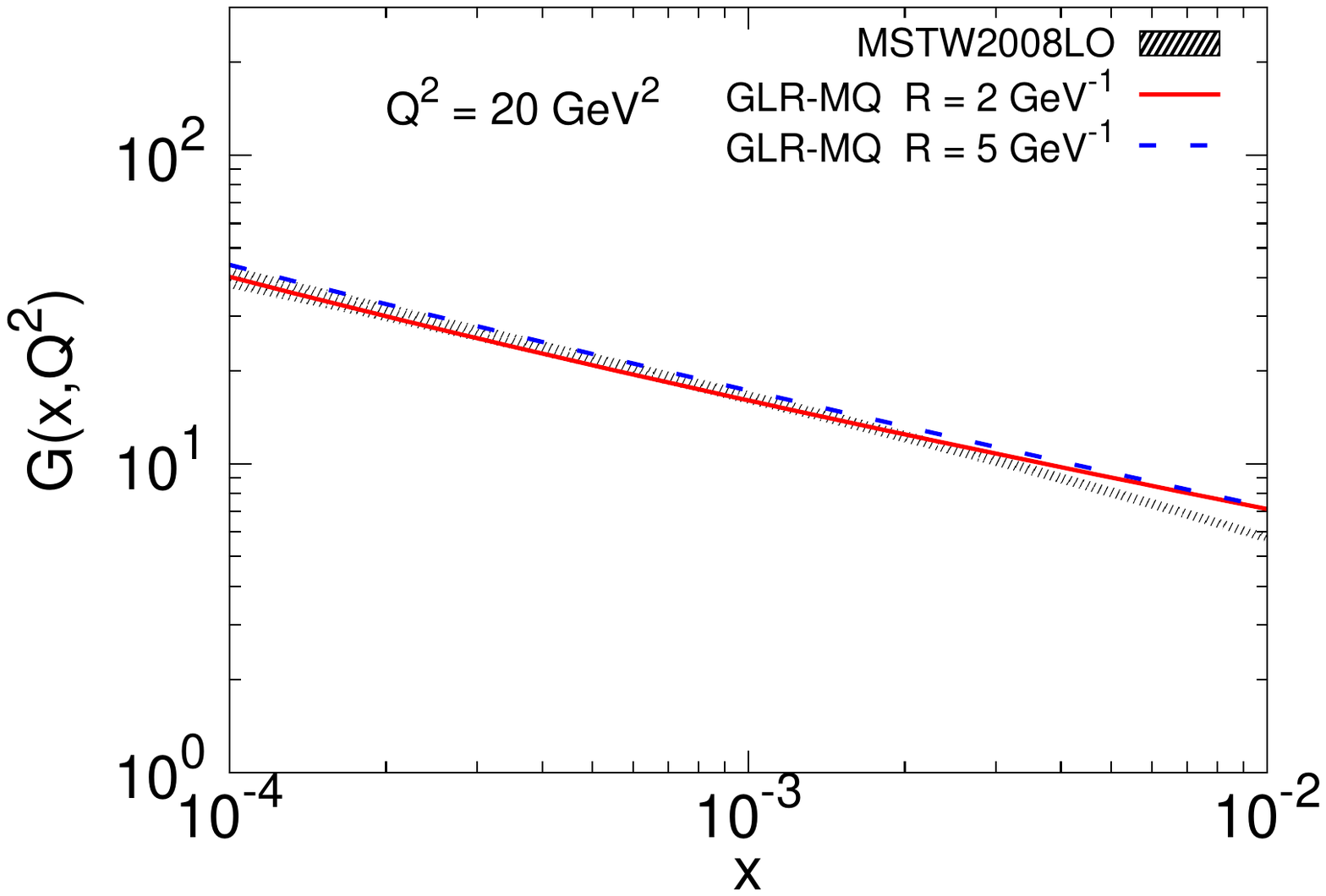}
%\vspace{0cm}\hspace{-0.3cm}
%%%%%%%%%%%%%%%%%%%%%%%%%%%%%%%%%%%%%%%%%%%%%%%%%%
\end{center}
\vspace{-0.5cm}
\caption{The gluon distribution functions $G(x,Q^2)$ as a function of Bjorken-$x$ at different virtuality $Q^{2}$. The solid lines represent the results from $R = 2~\text{GeV}^{-1}$, the dashed lines represent the results from $R = 5~\text{GeV}^{-1}$, and the bands represent the data from MSTW2008LO with error bars.}
\label{fig:G_x}
\end{figure}

Figure \ref{fig:G_Q2} shows the results of gluon distribution functions $G(x,Q^2)$ as a function of virtuality $Q^{2}$ at different Bjorken-$x$.  The solid lines represent the results from $R = 2~\text{GeV}^{-1}$, the dashed lines represent the results from $R = 5~\text{GeV}^{-1}$, and the bands represent the data from MSTW2008LO with error bars \cite{Martin:2009iq}.  As we have seen, our results from analytical solution are able to reproduce the behavior of the gluon distribution in most of the kinematic ranges we have considered. The agreement in the small $Q^2$ and very small-$x$ becomes worse as shown in the bottom panel of Fig.\ref{fig:G_Q2}. This is because the GLR-MQ equation becomes invalid as one goes into the deep saturation region. As discussed in Ref.\cite{kovchegov_levin_2012}, a higher order nonlinear corrections should be taken into account in this region.

Figure \ref{fig:G_x} shows the results of gluon distribution functions $G(x,Q^2)$ as a function of Bjorken-$x$ at different virtuality $Q^{2}$. We can clearly see that our results are consistent with the MSTW2008LO data. Similar to the $Q^{2}$ dependent case, the analytical solution is not accurate enough to reproduce the MSTW2008LO data as shown in the top-left panel in Fig.\ref{fig:G_x}. The deviation between the GLR-MQ and MSTW2008LO is mainly due to the kinematic range is beyond the validity of the GLR-MQ equation. As we know, the saturation of gluon is characterized by the saturation momentum $Q_s$, which increases with the decrease of $x$. Therefore, a relative large $Q_s$ is obtained at extremely small-$x$ and small $Q^{2}$ is much closer to the deep saturation region. From Figs.\ref{fig:G_Q2} and \ref{fig:G_x}, we can see that a steeper behavior of gluon distribution is found at $R = 5~\text{GeV}^{-1}$ as compared with $R = 2~\text{GeV}^{-1}$. This finding is consistent with the results in Regge approximation \cite{Devee:2014fna}.

\begin{figure}[t!]
\begin{center}
\vspace{-1.0cm}
\includegraphics[width=0.6\textwidth]{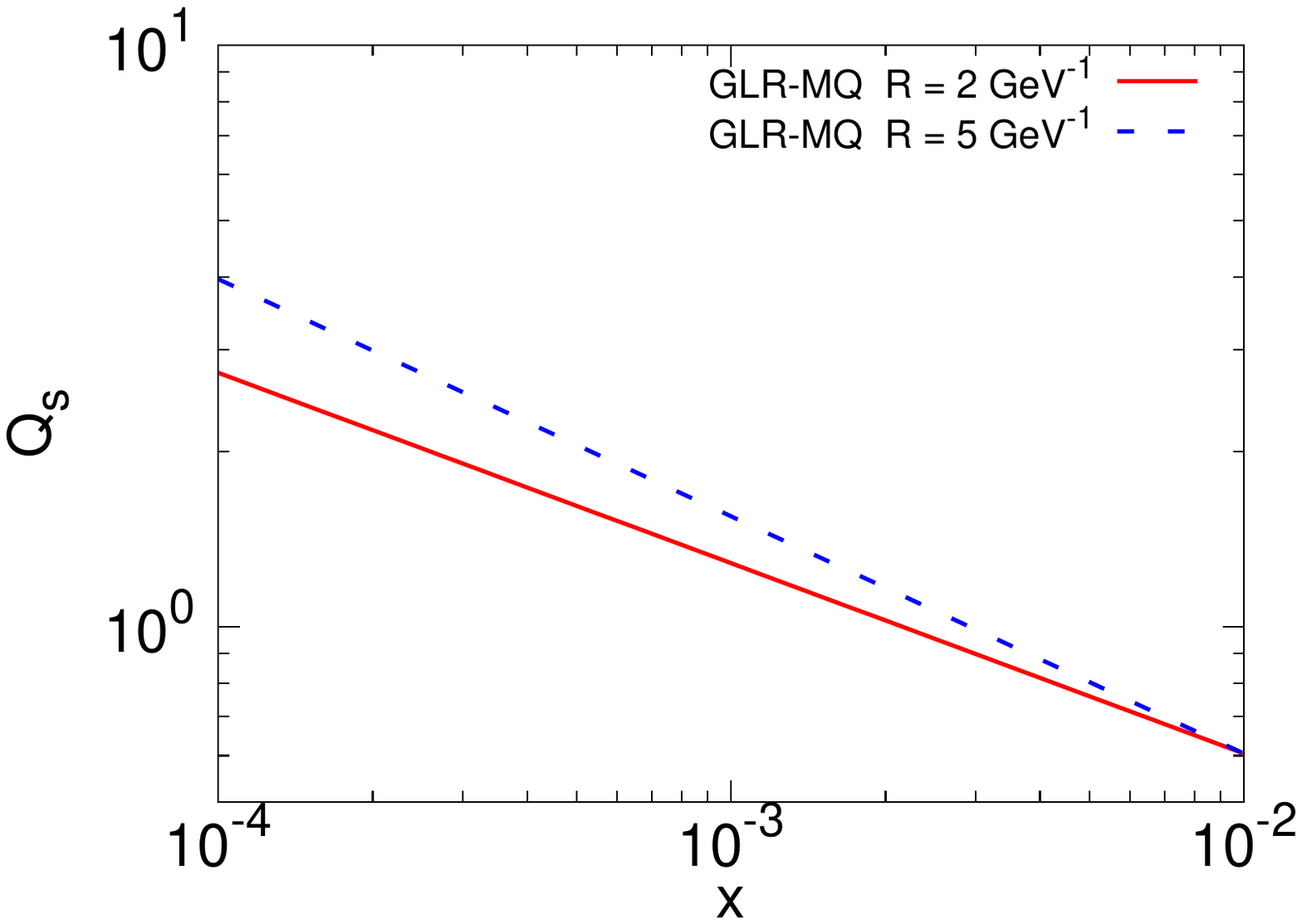}
%\vspace{-6.0cm}\hspace{-1.0cm}
\vspace{-0.5cm}
\end{center}
\caption{The saturation scale as a function of $x$. The solid lines represent the results at $R = 2~\text{GeV}^{-1}$ and the dashed lines represent the results at $R = 5~\text{GeV}^{-1}$.}
\label{fig:Qs}
\end{figure}

An important property of the GLR-MQ equation is that the nonlinear term will tame the growth of gluons and eventually leads to the saturation of gluons characterized by the saturation scale. The saturation scale can be estimated through defining the linear and nonlinear terms become equal or $\partial G/ \partial \ln Q^2 =0$ \cite{Eskola:2003gc}. In our analytical solution, the geometric scaling is an intrinsic property, which gives the saturation scale as
\be
\label{Qs}
Q_s^2(Y)\,=\,Q_0^2 e^{\frac{25\alpha Y}{6}}\ ,
\ee
where $Q_0$ is the initial saturation scale at $x=10^{-2}$.

Figure \ref{fig:Qs} shows the saturation scale at different correlation radii. For comparison, we adjust values of $Q_0$ to let saturation scale equal to than obtained from GBW model at $x=10^{-2}$ \cite{Golec-Biernat:1998zce}.
The saturation scale at $R = 2~\text{GeV}^{-1}$ and at $R = 5~\text{GeV}^{-1}$ are shown as the solid line and dash line, respectively. It is clear that the saturation momentum from our analytical solution obeys the exponential law $Q_s^2\,\propto\,Q_0^2 e^{\lambda Y}$, with $\lambda=0.657$ at $R = 2~\text{GeV}^{-1}$ and $\lambda=0.816$ at $R = 5~\text{GeV}^{-1}$. Thus the gluon distribution has a steeper growth at $R = 2~\text{GeV}^{-1}$. From Fig.\ref{fig:Qs}, we can see that $Q_s^2~\sim10~\text{GeV}^2$ at $x=10^{-4}$, which is considerably larger than $5~\text{GeV}^2$. So, the GLR-MQ becomes invalid at small $Q^2$ and very small-$x$, which support the finding in Figs.\ref{fig:G_Q2} and \ref{fig:G_x}.

\section{conclusions and discussions}
An exact analytical solution for the GLR-MQ equation at fixed coupling constant has been derived utilizing the homogeneous balance method. By fitting the MSTW2008LO gluon distribution function data, we obtain the definite solution of the GLR-MQ equation. The gluon distribution functions from our analytical solution show a good consistence with the MSTW2008LO data at $10^{-4} \leq x \leq 10^{-2}$ and $5~\text{GeV}^2 \leq Q^{2} \leq 20~\text{GeV}^2$, which indicates that our analytical solution from the homogeneous balance method is able to describe the behavior of gluon distribution. More importantly, the geometric scaling is an intrinsic property of our analytical solution. Based on the geometric scaling, we extract the saturation scale, which is found to obey an exponential law.

In this paper, we focus on the small-$x$ gluon behavior at fixed coupling constant. However, the running coupling and the quark composition can also affect the gluon behavior. Therefore, an analytical analysis of the effects of the nonlinear GLR-MQ equation with running coupling and the quark composition corrections is worth exploring in future studies. In addition, the investigation about the exclusive vector meson production within the saturation framework indicates that the proton has substructure (hot spot) \cite{Mantysaari:2016ykx}. Our analytical solution favors the gluon distribution concentration in the hot spot. So, further application of the gluon distributions from GLR-MQ is helpful to investigate the proton configuration.
%---------------------------------------------------------------------------

\begin{acknowledgments}
This work is supported by the Strategic Priority Research Program of Chinese Academy of Sciences under the Grant No. XDB34030301; Guangdong Major Project of Basic and Applied Basic Research Grant No. 2020B0301030008; National Natural Science Foundation of China under Grant No. 12165004; Guizhou Provincial Basic Research Program (Natural Science) under Grant No. QKHJC-ZK[2023]YB027; Education Department of Guizhou Province under Grant No. QJJ[2022]016.
\end{acknowledgments}

\bibliographystyle{JHEP-2modlong}
\bibliography{refs}

\end{document}